\documentclass[prl,showpacs,superscriptaddress,twocolumn,floatfix,nobalancelastpage]{revtex4}%
\usepackage{amsmath}
\usepackage{graphicx}
\usepackage{bm}
\usepackage{amsfonts}
\usepackage{amssymb}%

\begin{document}
\title{Symmetry Classes in Graphene Quantum Dots: Universal\\
Spectral Statistics, Weak Localization, and Conductance Fluctuations}

\author{J\"urgen Wurm}
\affiliation{Institut f\"ur Theoretische Physik, Universit\"at Regensburg, D-93040, Germany}
\affiliation{Department of Physics, Duke University, Box 90305,
Durham, North Carolina 27708-0305}

\author{Adam Rycerz}
\affiliation{Institut f\"ur Theoretische Physik, Universit\"at Regensburg, D-93040, Germany}
\affiliation{Marian Smoluchowski Institute of Physics, Jagiellonian University,
Reymonta 4, PL--30059 Krak\'{o}w, Poland}

\author{\.{I}nan\c{c} Adagideli}
\affiliation{Institut f\"ur Theoretische Physik, Universit\"at Regensburg, D-93040, Germany}

\author{Michael Wimmer}
\affiliation{Institut f\"ur Theoretische Physik, Universit\"at Regensburg, D-93040, Germany}

\author{Klaus Richter}
\affiliation{Institut f\"ur Theoretische Physik, Universit\"at Regensburg, D-93040, Germany}

\author{Harold U. Baranger}
\affiliation{Department of Physics, Duke University, Box 90305,
Durham, North Carolina 27708-0305}

\date{\today}


\begin{abstract}
We study the symmetry classes of graphene quantum dots, both open and closed, 
through the conductance and energy level statistics. For abrupt termination of the lattice, 
these properties are well described by the standard orthogonal and unitary ensembles. 
However, for smooth mass confinement, special time-reversal symmetries 
associated with the sublattice and valley degrees of freedom are
critical: they lead to block diagonal Hamiltonians and scattering matrices
with blocks belonging to the unitary symmetry class even at zero magnetic field. 
While the effect of this structure is clearly seen in the conductance of open dots, it 
is suppressed in the spectral statistics of closed dots, because the
intervalley scattering time is shorter than the time required to resolve a level spacing in the
closed systems but longer than the escape time of the open systems.
\end{abstract}

\maketitle

Single atomic layers of graphite, known as graphene, have attracted intense experimental and theoretical attention due to its unusual band structure
and hence exotic electronic properties \cite{Geim2007, Castro2008}.
Moreover, graphene's true two-dimensional nature and high mobility make it an attractive alternative for studying low dimensional electron systems such as quantum dots
\cite{Silvestrov2007, Ponomarenko2008, Stampfer2008, 
%Stampfer2008a, Guettinger2008, 
Stampfer2008b}. 
Recent experiments on the spectra of graphene quantum dots \cite{Ponomarenko2008} found evidence for a time reversal (TR) symmetry broken state in the absence of magnetic field, raising questions about the possible origin of such states.
Some time ago Berry and Mondragon \cite{Berry1987} proposed one such mechanism of TR symmetry breaking, namely infinite-mass confinement. 
In graphene dots, edge magnetization 
might produce such an effective mass term at the edges of the graphene flakes \cite{Fujita1996, Wimmer2008}, but whether this term is strong enough to change the universality class of the graphene quantum dots has not been established.

In this work we study universalities in the spectrum and conductance of graphene quantum dots in both the closed Coulomb blockade and the open ballistic regime, respectively. 
Universal properties are generally determined by the symmetries of the
Hamiltonian or the scattering matrix \cite{Mehta2004}.
Thus one expects that the universality class displayed by the spectrum of a closed quantum dot should be identical to that displayed by the conductance of a corresponding open dot. 
Here we show that this naive expectation is not true: the universality class of the conductance can be different from that of the spectrum.
The main reason behind this paradox is the separation of time scales characterizing the conductance (escape time) and the spectrum (Heisenberg time, i.e.\,inverse level spacing), allowing scattering times to be smaller than one but larger than the other.
To demonstrate this scenario, we first focus on closed graphene dots and show that their spectral statistics is described by the orthogonal symmetry class even in the presence of collinear edge magnetization, ruling out Berry and Mondragon's mechanism \cite{Berry1987} for TR symmetry breaking in this case. We next treat quantum transport through open dots and show that edge magnetism is enough to change the symmetry class, so that the conductance is described by the unitary ensemble.

\begin{figure}[tb]
\includegraphics[width=1\columnwidth]{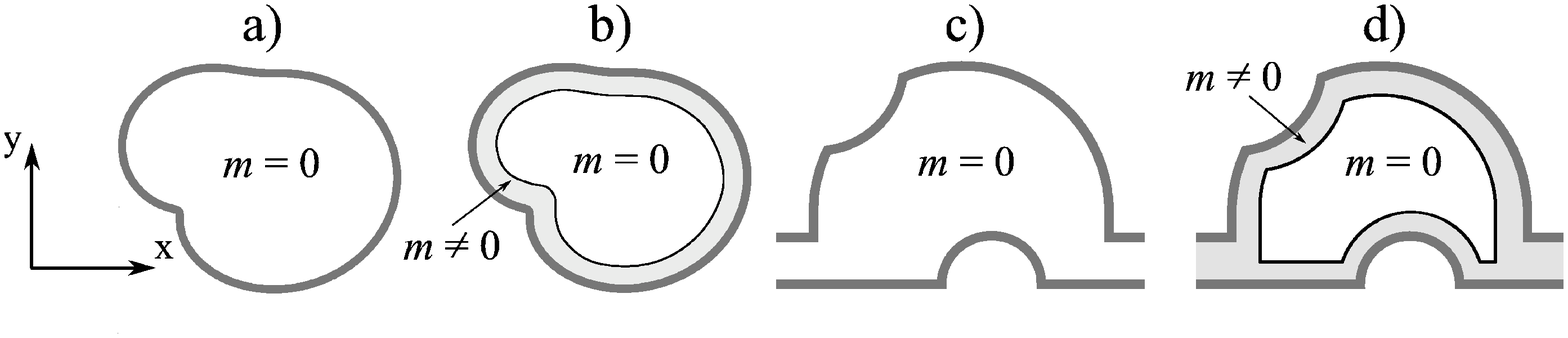}
\caption{Systems studied numerically (schematic). (a),(b) Africa billiard. (c),(d) Half-stadium with two identical leads; left-right symmetry is broken by cutting out circular segments at the top left and bottom right. The graphene lattice is terminated abruptly in (a) and (c), while smooth mass confinement is used in (b) and (d).}
\label{fig:Schemes}
\end{figure}

{\it Symmetries of the Hamiltonian---}The effective Hamiltonian for low energies and long length scales is the well-known Dirac Hamiltonian (the spin is omitted here),
\begin{equation}
\label{Heff}
\begin{split}
 & H_{\mathrm{eff}} = v (p_x-eA_x)\sigma_x\otimes\tau_z + v (p_y-eA_y)\sigma_y\otimes\tau_0 \\
 & \phantom{H_{\mathrm{eff}} = } + m(x,y) \sigma_z\otimes\tau_0 \;,
\end{split}
\end{equation}
where the Pauli matrices $\sigma_i$ and $\tau_i$ act on sublattice and valley degrees of freedom, respectively, and the index $i=0$ denotes the unit matrix. The boundary of the graphene flake is critical for its properties; we distinguish two physically relevant boundary types:
(i)~an abrupt termination of the graphene lattice, and (ii)~confinement by the mass term in Eq.\,(1). 
In the former case, $m(x,y)\equiv 0$; the boundary is disordered on the lattice scale and contains valley mixing armchair edges. In case (ii), while the lattice eventually terminates, the confinement is due to the smooth mass term which prevents the particles from feeling the rough boundary and thus suppresses the intervalley scattering. The mass term
may originate from an effective staggered potential caused by possible edge magnetization of graphene flakes \cite{Fujita1996, Wimmer2008}.

The symmetries of the problem are defined through three antiunitary operators \cite{Suzuura2002,Ostrovsky2007}: time reversal $\mathcal{T}$, and two ``special time reversal'' operators $\mathcal{T}_{\text{sl}}$ and $\mathcal{T}_{\text{v}}$, associated with 
either the \textit{sublattice} or \textit{valley} pseudospin:
\begin{equation}
  \mathcal{T} = (\sigma_0\otimes\tau_x)\mathcal{C},~~\mathcal{T}_\text{sl} = -i (\sigma_y\otimes\tau_0)\mathcal{C},~~\mathcal{T}_{\text{v}} = -i(\sigma_0\otimes\tau_y)\mathcal{C}.
\end{equation}
$\mathcal{C}$ denotes complex conjugation.
For abrupt termination, the two sublattices are inequivalent and boundary scattering mixes the valleys, so both special TR symmetries are irrelevant \cite{shortrangepotentials}.
For $B=0$, $\mathcal{T}$ commutes with $H_{\mathrm{eff}}$, leading to the orthogonal symmetry class. When $B \neq 0$, the Hamiltonian falls into the unitary ensemble.

For smooth mass confinement, intervalley scattering is small, so that the system largely consists of two independent subsystems, one for each valley. Each subsystem lacks TR symmetry, even at zero magnetic field, because $\mathcal{T}$ commutes only with the full $H_{\text{eff}}$, while $\mathcal{T}_{\text{sl}}$ is broken by the mass term. Thus, the Hamiltonian for a single valley corresponds to the unitary symmetry class. For zero magnetic field, however, $H_{\mathrm{eff}}$ commutes with $\mathcal{T}_{\text{v}}$ while $\mathcal{T}_{\text{v}}^2 = -I$. Kramers' theorem then guarantees the degeneracy of the eigenvalues of the full Hamiltonian \cite{Messiah1970}. Since the $\tau_y$ in $\mathcal{T}_{\text{v}}$ switches the valleys, the degenerate states do not lie in the same valley. Thus the Hamiltonian consists of two degenerate blocks with unitary symmetry. Upon applying a magnetic field, $H_{\mathrm{eff}}$ does not commute with $\mathcal{T}_{\text{v}}$,  and the valleys are no longer degenerate.

{\it Spectral statistics---}To exhibit the universality classes of closed 
graphene dots, we focus on the level spacing distribution for an Africa billiard \cite{Berry1987} with either abrupt termination or smooth mass confinement (Fig.\,\ref{fig:Schemes}). For the numerical work, we use the tight-binding Hamiltonian 
\begin{equation}
\label{eq:Htb}
 H_{\text{tb}} = \sum_{\langle i,j \rangle} t_{ij}\, c_i^{\dagger}c_j + \sum_{i} m_i\, c_i^{\dagger}c_i
\end{equation}
where $i$ and $j$ are nearest neighbors. The staggered potential $m_i=m(x_i,y_i)$, corresponding to a mass term, is positive (negative) if $i$ belongs to sublattice A (B). A magnetic field can be introduced via
$t_{ij} = -t\exp\left(i\frac{2\pi}{\Phi_0}\int_{\vec{r}_i}^{\vec{r}_j}\vec{A}\cdot d\vec{r}\right)$, with the flux quantum $\Phi_0~=~h/e$.
The lattice points are determined by cutting an Africa billiard out of a graphene plane [Fig.\,\ref{fig:Schemes}(a),(b)] 
with $x$ being a zigzag direction.
For smooth mass confinement, the mass term is zero in the interior but non-zero within a distance $W$ of the boundary [see Fig.\,\ref{fig:Schemes}(b)]; it starts from zero at the inner border of this region (black line in sketch) and increases quadratically: $ m(x,y)~=~\omega^2 \left[\delta(x,y)-W\right]^2/2$,
where $\delta(x,y)$ is the distance to the boundary and $\omega$ is a constant.

\begin{figure}[tb]
\includegraphics[width=\columnwidth]{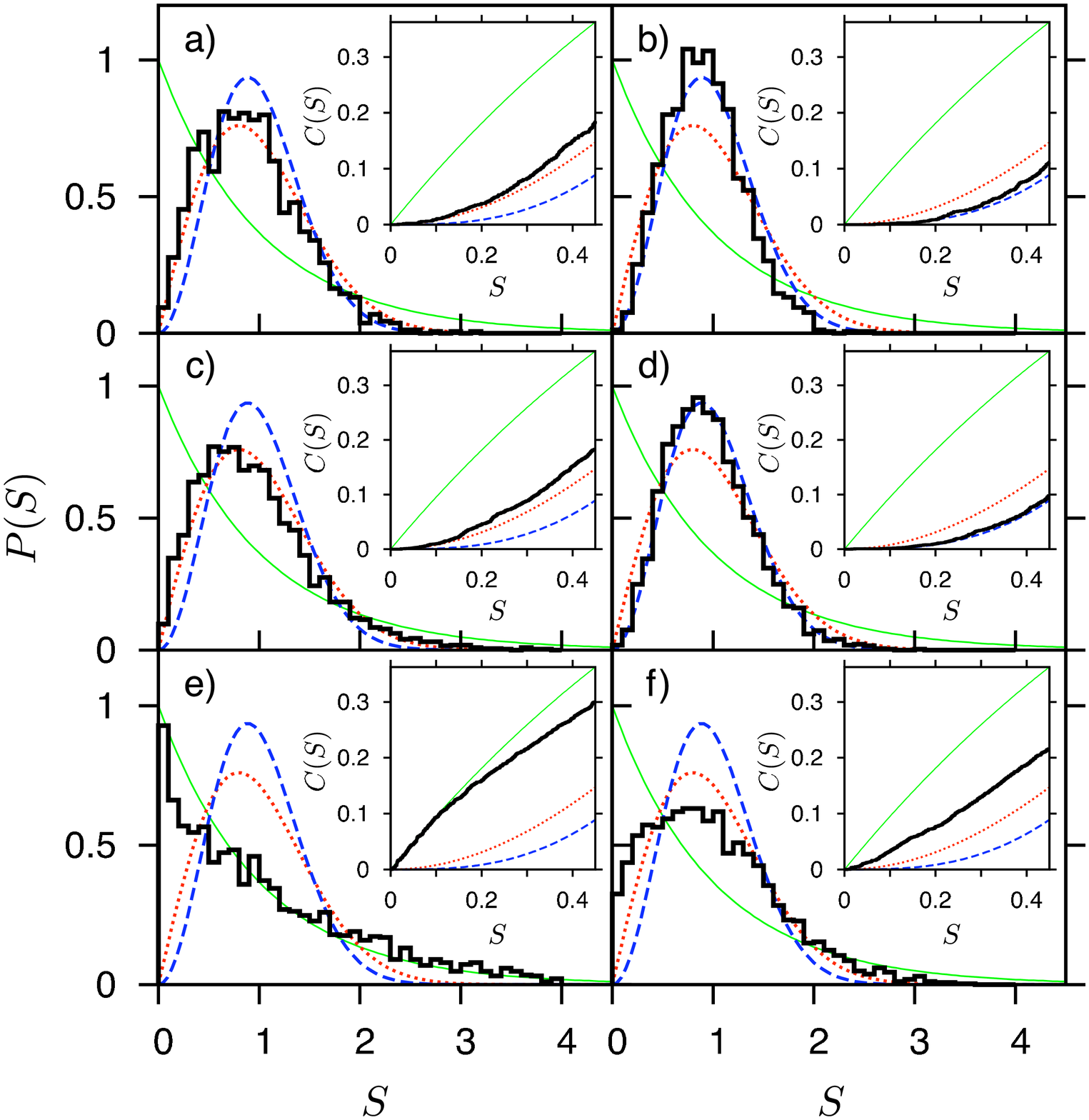}
\caption{(Color online)
Level-spacing distribution $P(S)$ for an Africa flake consisting of $68169$ carbon atoms using about 3000 energy levels in the range $[-0.5~t,0.5~t]$.
(a),(b) Abrupt lattice termination. 
(c),(d) Smooth mass confinement with $W=4.5\sqrt{3}a$, $\omega=0.15~\sqrt{t}/a$. 
(e),(f) Smooth mass confinement with $W=16.5\sqrt{3}a$, $\omega=0.041~\sqrt{t}/a$.
$a\approx 0.25~nm$ is the graphene lattice constant.
The \emph{left} panels are for $\Phi=0$, while the \emph{right} panels are for $\Phi=0.7\Phi_0$. Insets in each panel present the integrated distributions $C(S)=\int_0^SP(S')dS'$. Numerical results are shown with solid thick black lines, whereas the thin lines are for Poisson (green solid), GOE (red dotted), and GUE (blue dashed) statistics. 
}
\label{fig:africa}%
\end{figure}

Fig.\,\ref{fig:africa} shows the level spacing distribution for both abrupt termination and smooth confinement in an Africa graphene dot. For abrupt termination (top panels in Fig.\,\ref{fig:africa}), the statistics are consistent with the Gaussian \textit{orthogonal} ensemble (GOE) when $B=0$ and with the Gaussian \textit{unitary} ensemble (GUE) upon introduction of a magnetic field. This is expected from the symmetry considerations above. 

For smooth mass confinement, the results are surprising: the statistics are \textit{not} the expected GUE but rather are GOE for large systems [Fig.\,\ref{fig:africa}(c)] with a crossover to Poisson for smaller systems [Fig.\,\ref{fig:africa}(e)]. 
This crossover reflects the role of localized edge states present for energies near the Dirac point which follow Poisson statistics. Edge states dominate in small systems, but for larger systems their spectral weight diminishes, giving rise to the crossover to GOE statistics. We believe this is why the numerical level statistics in Ref.~\cite{Raedt2008} does not fit well to either Poisson or Gaussian ensembles. 

The reason that we find orthogonal rather than unitary statistics for a large dot is more subtle: Though our mass confinement is fairly smooth, there is some residual intervalley scattering. If the intervalley scattering time is shorter than the relevant time scale for the level spacing (i.e.\ the Heisenberg time), time-reversal symmetry will be restored. To probe this idea further, we consider another observable with a very different time scale, namely the conductance of an open cavity for which the time scale is the escape time.

\begin{figure}[ptb]
\includegraphics[width=0.96\columnwidth]{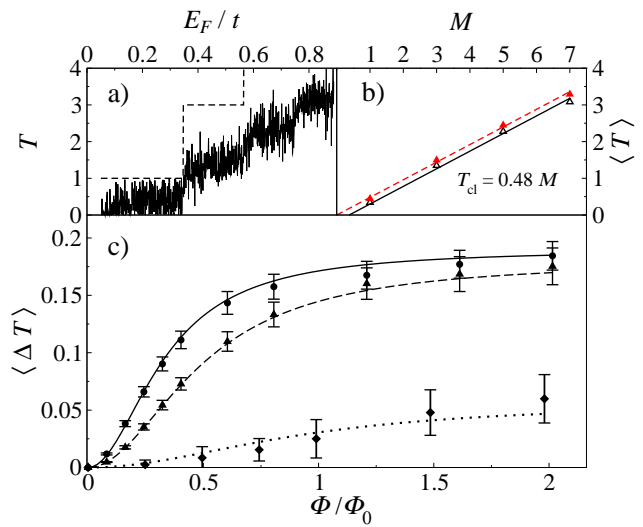}
\caption{(Color online) Average conductance: weak localization.
(a) Transmission as a function of energy for an abruptly terminated billiard [Fig.\,\ref{fig:Schemes}(c)] with zigzag leads (solid line). The dashed line shows the number of open channels in the leads, $M$. (b) Average transmission as a function of $M$ for the same system with $\Phi=0$ (solid black line, open triangles) and $\Phi=1.6~\Phi_0$ (dashed red line, full triangles). (c) Change in the average transmission as a function of the magnetic flux. Circles: Abrupt termination with armchair leads (1-7 open channels). The fit (solid line) yields $A_0 = 1.5 A_B$ and $\mathcal{R}=0.19$. Triangles: Abrupt termination with zigzag leads (3-7 open channels). The fit (dashed line) yields $A_0 = 1.0 A_B$ and $\mathcal{R}=0.18$. Diamonds: Smooth mass confinement [Fig.\,\ref{fig:Schemes}(d)] (2-8 open channels). The fit (dotted line) yields $A_0 = 0.54 A_B$ and $\mathcal{R}=0.057$ (Parameters of the billiards given in \cite{weaklocparams}).
}
\label{fig:WL}
\end{figure}

{\it Quantum transport: weak localization---}The conductance of a cavity attached to two leads [Fig.\,\ref{fig:Schemes}(c),(d)] is proportional to the quantum mechanical transmission probability from one lead to the other. We use a recursive Green function method \cite{Wimmer2008a} to find the transmission for tight-binding cavities with either abrupt or smooth boundaries. First, we focus on the average transmission $\langle T\,\rangle$, where the average is performed with respect to the Fermi energy -- see Figs.\,\ref{fig:WL}(a) and (b) for an example of $T(E_F)$ and its average. We find that $\langle T\,\rangle \approx 0.5 M$ to leading order in the number of open channels in the leads $M$ [Fig.\,\ref{fig:WL}(b)]: particles are transmitted or reflected with about equal probability. The next order correction, known as the weak localization correction, is the y-axis intercept in Fig.\,\ref{fig:WL}(b). It has been studied theoretically for diffusive graphene systems \cite{McCann2006, Khveshchenko2006}. As expected for abrupt termination, there is no offset for large enough magnetic fields. 

We now focus on the average magnetoconductance of our graphene billiards to study the weak localization correction in more detail. We compare the magnetoconductance data to the semiclassical Lorentzian prediction \cite{Baranger1993}:
\begin{equation}
\label{eq:DT(B)}
 \langle\Delta T(B)\,\rangle  \equiv \langle T(B)-T(0) \rangle = \mathcal{R}/[1+(\Phi_0/2A_0 B)^2]
\end{equation}
where $\mathcal{R}$ is the total magnitude of the effect, $A_0$ is the typical area enclosed by classical paths. According to random matrix theory (RMT) \cite{Baranger1994, Jalabert1994}, $\mathcal{R}=M/(4M +2)$ is the difference between the average conductance in systems with unitary and orthogonal symmetry (weak localization is suppressed for unitary symmetry), in agreement with the semiclassical theory~\cite{Richter2002} for large $M$.

The numerical results obtained by averaging over an energy window are in good agreement with Eq.\,(\ref{eq:DT(B)}) [Fig.\,\ref{fig:WL}(c)]. The fit parameter $A_0$ is of the order of the billiard area $A_B$ so that weak localization is suppressed for a magnetic flux of about $\Phi_0$. For the abruptly terminated billiard with armchair leads, we find $\mathcal{R}=0.19$ while the corresponding RMT value is $\overline{\mathcal{R}}_{\mathrm{RMT}} = 0.20$. For zigzag leads (in the multi-mode regime), we find $\mathcal{R}=0.18$ while $\overline{\mathcal{R}}_{\rm{RMT}} = 0.22$. Thus, for the abruptly terminated billiards, our numerical results agree with RMT for the expected symmetry classes. 

For smooth mass confinement, the expected symmetry classes are unitary, both in the absence and presence of a magnetic field. Thus, no weak localization correction is expected. Numerically, a very small weak localization correction is visible: $\mathcal{R}=0.057$. We assign the slight increase of $ \langle\Delta T\,\rangle $
%in Fig.\,\ref{fig:WL} 
to weak residual intervalley scattering. 

{\it Conductance fluctuations---}To show the change in symmetry class upon applying a magnetic field for smooth mass confinement, we turn to conductance fluctuations. Universal conductance fluctuations for the orthogonal symmetry class were found in transport calculations on weakly-disordered, rectangular graphene samples with zigzag edges \cite{Rycerz2007}. Here, to obtain direct information about the symmetry classes, we investigate the magnitude of the conductance fluctuations in chaotic cavities as a function of energy. 
The RMT results for the variance of the conductance as a function of $M$ are given in \cite{Baranger1994} [Eq.\,3(b)] and \cite{Jalabert1994} (Eq.\,11), for the cases of the
circular \textit{orthogonal} (COE) and the circular \textit{unitary} (CUE) ensemble.

\begin{figure}
\includegraphics[width=0.96\columnwidth]{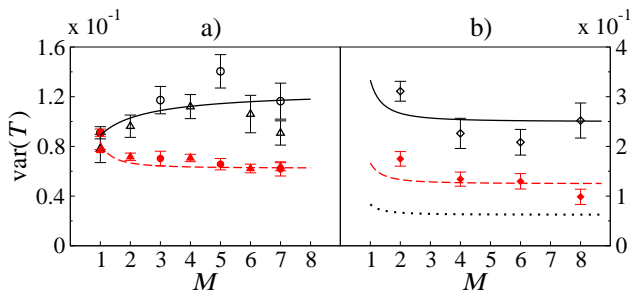}
\caption{(Color online) 
Conductance fluctuations: 
Variance of the transmission as a function of the number of open channels in the leads (same cavities as in Fig.\,\ref{fig:WL}). 
(a) Abruptly terminated boundary. $B=0$ (black open symbols) and $B \neq 0$ (red full symbols) results are in good agreement with the corresponding RMT values, orthogonal (COE, black solid line) and unitary (CUE, red dashed line). The unitary data uses several values for the magnetic field in the range $\Phi \in [0.8,2.4]\Phi_0$; both armchair leads (triangles) and zigzag leads (circles) are used. 
(b) Smooth mass confinement. Zero field (black open symbols) and $\Phi = 2.0 \,\Phi_0$ (red full symbols) results are compared to 1, 2, and 4 times the CUE values (black dotted, red dashed, and black solid lines). 
}
\label{fig:VarG}
\end{figure}

In Fig.\,\ref{fig:VarG} we present the numerical results for the conductance fluctuations. For the cavities with abruptly terminated edges, $\mathrm{var} (T)$ clearly agrees with the COE result when $B=0$, while it follows the CUE curve if a magnetic field is present. This is as expected from the symmetry considerations and weak localization results. 

For smooth mass confinement, Fig.\,\ref{fig:VarG}(b) shows that the magnitude of the fluctuations at zero magnetic field is much larger than the COE or CUE values. Rather, it is approximately four times the CUE value. When a magnetic field is applied, $\mathrm{var} (T)$ becomes smaller, about twice the CUE value. This is consistent with the symmetry considerations given at the beginning of this paper: An ensemble of transmission matrices each with two identical blocks implies that $\mathrm{var} (T)$ will be 4 times the value for a single block. However, an ensemble of transmission matrices, each with two uncorrelated blocks, yields the sum of the single blocks' values. Since the blocks are expected to be unitary in the case of smooth mass confinement, with or without a magnetic field, the result in Fig.\,\ref{fig:VarG}(b) follows.

To summarize, dots formed by mass confinement do \emph{not} follow expectations derived from the effective Dirac equation. While the transmission statistics follow from the expected block unitary structure, the spectral statistics show orthogonal or even Poisson statistics. Thus, the spectral and transmission statistics follow different ensembles! This paradox arises from residual intervalley scattering in our system -- though the confinement used, $m(x,y)$, varies on a scale of 10-30 lattice constants for our dots, some weak lattice effects always remain. The time scale appropriate for transmission statistics is the escape time from the cavity while the time scale for spectral statistics is the much longer inverse level spacing. Hence if the intervalley scattering time lies between the two, different behavior can result. Our study suggests that it will be more fruitful to look for smooth confinement effects, such as the Berry and Mondragon breaking of orthogonal symmetry without a magnetic field \cite{Berry1987}, in open rather than closed dots.

We thank Denis Ullmo and Eduardo Mucciolo for helpful discussions.
The work at Duke was supported in part by the NSF (Grant No. DMR-0506953) and by the DAAD. 
A.R. acknowledges support from the Alexander von Humboldt foundation and the Polish
Ministry of Science (Grant No. 1--P03B--001--29).
We further acknowledge support by the DFG (through SFB 689). 

After this work was completed, we became aware of a preprint on spectral statistics in nanotube-like structures, Ref.\,\onlinecite{Amanatidis2008}.


\vspace*{-0.2in}
\begin{thebibliography}{26}
\expandafter\ifx\csname natexlab\endcsname\relax\def\natexlab#1{#1}\fi
\expandafter\ifx\csname bibnamefont\endcsname\relax
  \def\bibnamefont#1{#1}\fi
\expandafter\ifx\csname bibfnamefont\endcsname\relax
  \def\bibfnamefont#1{#1}\fi
\expandafter\ifx\csname citenamefont\endcsname\relax
  \def\citenamefont#1{#1}\fi
\expandafter\ifx\csname url\endcsname\relax
  \def\url#1{\texttt{#1}}\fi
\expandafter\ifx\csname urlprefix\endcsname\relax\def\urlprefix{URL }\fi
\providecommand{\bibinfo}[2]{#2}
\providecommand{\eprint}[2][]{\url{#2}}


\bibitem[{\citenamefont{Geim and Novoselov}(2007)}]{Geim2007}
\bibinfo{author}{\bibfnamefont{A.~K.} \bibnamefont{Geim}} \bibnamefont{and}
  \bibinfo{author}{\bibfnamefont{K.~S.} \bibnamefont{Novoselov}},
  \bibinfo{journal}{Nature Materials} \textbf{\bibinfo{volume}{6}},
  \bibinfo{pages}{183 } (\bibinfo{year}{2007}).

\bibitem[{\citenamefont{Castro~Neto et~al.}(2008)\citenamefont{Castro~Neto,
  Guinea, Peres, Novoselov, and Geim}}]{Castro2008}
\bibinfo{author}{\bibfnamefont{A.~H.} \bibnamefont{Castro~Neto}},
  \bibinfo{author}{\bibfnamefont{F.}~\bibnamefont{Guinea}},
  \bibinfo{author}{\bibfnamefont{N.~M.~R.} \bibnamefont{Peres}},
  \bibinfo{author}{\bibfnamefont{K.~S.} \bibnamefont{Novoselov}},
  \bibnamefont{and} \bibinfo{author}{\bibfnamefont{A.~K.} \bibnamefont{Geim}},
  \bibinfo{journal}{arXiv:0709.1163 (to be published in Rev. Mod. Phys.)}  (\bibinfo{year}{2008}).

\bibitem[{\citenamefont{Silvestrov and Efetov}(2007)}]{Silvestrov2007}
\bibinfo{author}{\bibfnamefont{P.~G.} \bibnamefont{Silvestrov}}
  \bibnamefont{and} \bibinfo{author}{\bibfnamefont{K.~B.}
  \bibnamefont{Efetov}}, \bibinfo{journal}{Phy. Rev. Lett.}
  \textbf{\bibinfo{volume}{98}}, \bibinfo{eid}{016802} (\bibinfo{year}{2007}).

\bibitem[{\citenamefont{Ponomarenko et~al.}(2008)\citenamefont{Ponomarenko,
  Schedin, Katsnelson, Yang, Hill, Novoselov, and Geim}}]{Ponomarenko2008}
\bibinfo{author}{\bibfnamefont{L.~A.} \bibnamefont{Ponomarenko}},
  \bibinfo{author}{\bibfnamefont{F.}~\bibnamefont{Schedin}},
  \bibinfo{author}{\bibfnamefont{M.~I.} \bibnamefont{Katsnelson}},
  \bibinfo{author}{\bibfnamefont{R.}~\bibnamefont{Yang}},
  \bibinfo{author}{\bibfnamefont{E.~W.} \bibnamefont{Hill}},
  \bibinfo{author}{\bibfnamefont{K.~S.} \bibnamefont{Novoselov}},
  \bibnamefont{and} \bibinfo{author}{\bibfnamefont{A.~K.} \bibnamefont{Geim}},
  \bibinfo{journal}{Science} \textbf{\bibinfo{volume}{320}},
  \bibinfo{pages}{356} (\bibinfo{year}{2008}).

\bibitem[{\citenamefont{Stampfer
  et~al.}(2008{\natexlab{a}})\citenamefont{Stampfer, Guettinger, Molitor, Graf,
  Ihn, and Ensslin}}]{Stampfer2008}
\bibinfo{author}{\bibfnamefont{C.}~\bibnamefont{Stampfer}},
  \bibinfo{author}{\bibfnamefont{J.}~\bibnamefont{Guettinger}},
  \bibinfo{author}{\bibfnamefont{F.}~\bibnamefont{Molitor}},
  \bibinfo{author}{\bibfnamefont{D.}~\bibnamefont{Graf}},
  \bibinfo{author}{\bibfnamefont{T.}~\bibnamefont{Ihn}}, \bibnamefont{and}
  \bibinfo{author}{\bibfnamefont{K.}~\bibnamefont{Ensslin}},
  \bibinfo{journal}{Appl. Phys. Lett.} \textbf{\bibinfo{volume}{92}},
  \bibinfo{pages}{012102} (\bibinfo{year}{2008}{\natexlab{a}}).

\bibitem[{\citenamefont{Stampfer
  et~al.}(2008{\natexlab{c}})\citenamefont{Stampfer, Schnez, Guettinger,
  Hellmueller, Molitor, Shorubalko, Ihn, and Ensslin}}]{Stampfer2008b}
\bibinfo{author}{\bibfnamefont{C.}~\bibnamefont{Stampfer}},
  \bibinfo{author}{\bibfnamefont{S.}~\bibnamefont{Schnez}},
  \bibinfo{author}{\bibfnamefont{J.}~\bibnamefont{Guettinger}},
  \bibinfo{author}{\bibfnamefont{S.}~\bibnamefont{Hellmueller}},
  \bibinfo{author}{\bibfnamefont{F.}~\bibnamefont{Molitor}},
  \bibinfo{author}{\bibfnamefont{I.}~\bibnamefont{Shorubalko}},
  \bibinfo{author}{\bibfnamefont{T.}~\bibnamefont{Ihn}}, \bibnamefont{and}
  \bibinfo{author}{\bibfnamefont{K.}~\bibnamefont{Ensslin}},
  \bibinfo{journal}{arXiv:0807.2710}  (\bibinfo{year}{2008}{\natexlab{c}}).

\bibitem[{\citenamefont{Berry and Mondragon}(1987)}]{Berry1987}
\bibinfo{author}{\bibfnamefont{M.~V.} \bibnamefont{Berry}} \bibnamefont{and}
  \bibinfo{author}{\bibfnamefont{R.~J.} \bibnamefont{Mondragon}},
  \bibinfo{journal}{Proc. R. Soc. Lond. A} \textbf{\bibinfo{volume}{412}},
  \bibinfo{pages}{53} (\bibinfo{year}{1987}).

\bibitem[{\citenamefont{Fujita et~al.}(1996)\citenamefont{Fujita, Wakabayashi,
  Nakada, and Kusakabe}}]{Fujita1996}
\bibinfo{author}{\bibfnamefont{M.}~\bibnamefont{Fujita}},
  \bibinfo{author}{\bibfnamefont{K.}~\bibnamefont{Wakabayashi}},
  \bibinfo{author}{\bibfnamefont{K.}~\bibnamefont{Nakada}}, \bibnamefont{and}
  \bibinfo{author}{\bibfnamefont{K.}~\bibnamefont{Kusakabe}},
  \bibinfo{journal}{J. Phys. Soc. Jpn.} \textbf{\bibinfo{volume}{65}},
  \bibinfo{pages}{1920} (\bibinfo{year}{1996}).

\bibitem[{\citenamefont{Wimmer et~al.}(2008)\citenamefont{Wimmer, \.{I}nan\c{c}
  Adagideli, Berber, Tom\'{a}nek, and Richter}}]{Wimmer2008}
\bibinfo{author}{\bibfnamefont{M.}~\bibnamefont{Wimmer}},
  \bibinfo{author}{\bibnamefont{\.{I}nan\c{c} Adagideli}},
  \bibinfo{author}{\bibfnamefont{S.}~\bibnamefont{Berber}},
  \bibinfo{author}{\bibfnamefont{D.}~\bibnamefont{Tom\'{a}nek}},
  \bibnamefont{and} \bibinfo{author}{\bibfnamefont{K.}~\bibnamefont{Richter}},
  \bibinfo{journal}{Phys. Rev. Lett.} \textbf{\bibinfo{volume}{100}},
  \bibinfo{eid}{177207} (\bibinfo{year}{2008}).

\bibitem[{\citenamefont{Mehta}(2004)}]{Mehta2004}
\bibinfo{author}{\bibfnamefont{M.~L.} \bibnamefont{Mehta}},
  \emph{\bibinfo{title}{Random Matrices}} (\bibinfo{publisher}{Elsevier},
  \bibinfo{year}{2004}).

\bibitem[{\citenamefont{Suzuura and Ando}(2002)}]{Suzuura2002}
\bibinfo{author}{\bibfnamefont{H.}~\bibnamefont{Suzuura}} \bibnamefont{and}
  \bibinfo{author}{\bibfnamefont{T.}~\bibnamefont{Ando}},
  \bibinfo{journal}{Phys. Rev. Lett.} \textbf{\bibinfo{volume}{89}},
  \bibinfo{pages}{266603} (\bibinfo{year}{2002}).

\bibitem[{\citenamefont{Ostrovsky et~al.}(2007)\citenamefont{Ostrovsky, Gornyi,
  and Mirlin}}]{Ostrovsky2007}
\bibinfo{author}{\bibfnamefont{P.~M.} \bibnamefont{Ostrovsky}},
  \bibinfo{author}{\bibfnamefont{I.~V.} \bibnamefont{Gornyi}},
  \bibnamefont{and} \bibinfo{author}{\bibfnamefont{A.~D.}
  \bibnamefont{Mirlin}}, \bibinfo{journal}{Eur. Phys. J. ST}
  \textbf{\bibinfo{volume}{148}}, \bibinfo{pages}{63} (\bibinfo{year}{2007}).

\bibitem{shortrangepotentials}
 It has been shown \cite{Suzuura2002}, that short range potentials in general break the symmetry given by $\mathcal{T}_{\text{sl}}$.

\bibitem[{\citenamefont{Messiah}(1970)}]{Messiah1970}
\bibinfo{author}{\bibfnamefont{A.}~\bibnamefont{Messiah}},
  \emph{\bibinfo{title}{Quantum Mechanics}, vol. 2}
  (\bibinfo{publisher}{North-Holland}, Amsterdam, \bibinfo{year}{1970})\bibinfo{pages}{pp. 669-675}.

\bibitem[{\citenamefont{Raedt and Katsnelson}(2008)}]{Raedt2008}
\bibinfo{author}{\bibfnamefont{H.~D.} \bibnamefont{Raedt}} \bibnamefont{and}
  \bibinfo{author}{\bibfnamefont{M.}~\bibnamefont{Katsnelson}},
  \bibinfo{journal}{arXiv:0804.2758}  (\bibinfo{year}{2008}).

\bibitem[{\citenamefont{Wimmer and Richter}(2008)}]{Wimmer2008a}
\bibinfo{author}{\bibfnamefont{M.}~\bibnamefont{Wimmer}} \bibnamefont{and}
  \bibinfo{author}{\bibfnamefont{K.}~\bibnamefont{Richter}},
  \bibinfo{journal}{arXiv:0806.2739}  (\bibinfo{year}{2008}).

 \bibitem{weaklocparams}
Parameters of the open systems studied numerically:
(1) Abrupt termination with armchair leads: $A_B = (166a)^2$ and $E_F \in [0.08,0.84]$ (1-7 channels).
(2) Abrupt termination with zigzag leads: $A_B = (166a)^2$ and average taken using $E_F \in [0.35,0.89]$ (3-7 channels in the leads).
(3) Smooth mass confinement: $\omega=0.050~\sqrt{t}/a$, $W=20~a$ $A_B = (184a)^2$, and $E_F \in [0.07,0.45]$ (2-8 channels).

\bibitem[{\citenamefont{McCann et~al.}(2006)\citenamefont{McCann, Kechedzhi,
  Fal'ko, Suzuura, Ando, and Altshuler}}]{McCann2006}
\bibinfo{author}{\bibfnamefont{E.}~\bibnamefont{McCann}},
  \bibinfo{author}{\bibfnamefont{K.}~\bibnamefont{Kechedzhi}},
  \bibinfo{author}{\bibfnamefont{V.~I.} \bibnamefont{Fal'ko}},
  \bibinfo{author}{\bibfnamefont{H.}~\bibnamefont{Suzuura}},
  \bibinfo{author}{\bibfnamefont{T.}~\bibnamefont{Ando}}, \bibnamefont{and}
  \bibinfo{author}{\bibfnamefont{B.~L.} \bibnamefont{Altshuler}},
  \bibinfo{journal}{Phy. Rev. Lett.} \textbf{\bibinfo{volume}{97}},
  \bibinfo{eid}{146805} (\bibinfo{year}{2006}).

\bibitem[{\citenamefont{Khveshchenko}(2006)}]{Khveshchenko2006}
\bibinfo{author}{\bibfnamefont{D.~V.} \bibnamefont{Khveshchenko}},
  \bibinfo{journal}{Phy. Rev. Lett.} \textbf{\bibinfo{volume}{97}},
  \bibinfo{eid}{036802} (\bibinfo{year}{2006}).

\bibitem[{\citenamefont{Baranger et~al.}(1993)\citenamefont{Baranger, Jalabert,
  and Stone}}]{Baranger1993}
\bibinfo{author}{\bibfnamefont{H.~U.} \bibnamefont{Baranger}},
  \bibinfo{author}{\bibfnamefont{R.~A.} \bibnamefont{Jalabert}},
  \bibnamefont{and} \bibinfo{author}{\bibfnamefont{A.~D.} \bibnamefont{Stone}},
  \bibinfo{journal}{Phys. Rev. Lett.} \textbf{\bibinfo{volume}{70}},
  \bibinfo{pages}{3876} (\bibinfo{year}{1993});
  \bibinfo{journal}{Chaos} \textbf{\bibinfo{volume}{3}}, \bibinfo{pages}{665}
  (\bibinfo{year}{1993}).

\bibitem[{\citenamefont{Baranger and Mello}(1994)}]{Baranger1994}
\bibinfo{author}{\bibfnamefont{H.~U.} \bibnamefont{Baranger}} \bibnamefont{and}
  \bibinfo{author}{\bibfnamefont{P.~A.} \bibnamefont{Mello}},
  \bibinfo{journal}{Phys. Rev. Lett.} \textbf{\bibinfo{volume}{73}},
  \bibinfo{pages}{142} (\bibinfo{year}{1994}).

\bibitem[{\citenamefont{Jalabert et~al.}(1994)\citenamefont{Jalabert, Pichard,
  and Beenakker}}]{Jalabert1994}
\bibinfo{author}{\bibfnamefont{R.~A.} \bibnamefont{Jalabert}},
  \bibinfo{author}{\bibfnamefont{J.-L.} \bibnamefont{Pichard}},
  \bibnamefont{and} \bibinfo{author}{\bibfnamefont{C.~W.~J.}
  \bibnamefont{Beenakker}}, \bibinfo{journal}{Europhys. Lett.}
  \textbf{\bibinfo{volume}{27}}, \bibinfo{pages}{255} (\bibinfo{year}{1994}).

\bibitem[{\citenamefont{Richter and Sieber}(2002)}]{Richter2002}
\bibinfo{author}{\bibfnamefont{K.}~\bibnamefont{Richter}} \bibnamefont{and}
  \bibinfo{author}{\bibfnamefont{M.}~\bibnamefont{Sieber}},
  \bibinfo{journal}{Phys. Rev. Lett.} \textbf{\bibinfo{volume}{89}},
  \bibinfo{pages}{206801} (\bibinfo{year}{2002}).

\bibitem[{\citenamefont{Rycerz et~al.}(2007)\citenamefont{Rycerz, Tworzydlo,
  and Beenakker}}]{Rycerz2007}
\bibinfo{author}{\bibfnamefont{A.}~\bibnamefont{Rycerz}},
  \bibinfo{author}{\bibfnamefont{J.}~\bibnamefont{Tworzyd{\l}o}},
  \bibnamefont{and} \bibinfo{author}{\bibfnamefont{C.~W.~J.}
  \bibnamefont{Beenakker}}, \bibinfo{journal}{Europhys. Lett.}
  \textbf{\bibinfo{volume}{79}}, \bibinfo{pages}{57003} (\bibinfo{year}{2007}).

\bibitem[{\citenamefont{Amanatidis and Evangelou}(2008)}]{Amanatidis2008}
\bibinfo{author}{\bibfnamefont{I.}~\bibnamefont{Amanatidis}} \bibnamefont{and}
  \bibinfo{author}{\bibfnamefont{S.}~\bibnamefont{Evangelou}},
  \bibinfo{journal}{arXiv:0806.4884}  (\bibinfo{year}{2008}).


\end{thebibliography}
\end{document}